\documentclass[epj,referee]{svjour}

\usepackage{graphicx}

\begin{document}

\title{Study of the Ce(Rh$_{1-x}$Pd$_x$)$_2$Si$_2$ alloy:
evidence for itinerant character of the magnetic order in CeRh$_2$Si$_2$.}

\author{M. G\'omez Berisso\inst{1} \and P. Pedrazzini\inst{1} \and
J.G. Sereni\inst{1} \and O. Trovarelli\inst{2} \and C. Geibel\inst{2}
\and F. Steglich\inst{2}}

\institute{Lab. Bajas Temperaturas, Centro At\'omico Bariloche - IB
(CNEA) and CONICET, 8400 S.C. de Bariloche, Argentina \and Max-Planck
Institut for Chemical Physics of Solids, N\"othnitzer Str. 40,
D-01187 Dresden, Germany}

\date{\today}

\abstract{ We present electrical resistivity ($\rho(T)$) and specific
heat ($C_{\rm el}(T)$) measurements of alloys on the Rh rich side of
the phase diagram of the Ce(Rh$_{1-x}$Pd$_x$)$_2$Si$_2$ system and
compare the results with those obtained at intermediate and low Rh
concentrations. The analysis of the $x$-evolution of the entropy  and
the scaling behaviour of $C_{\rm el}(T)$ and $\rho(T)$ clearly
confirm a separation of the magnetic phase diagram into two regions.
The region $x \le 0.3$, showing a concentration independent
characteristic temperature $T_{\rm o} \approx  45 \,{\rm K}$, while
for $x > 0.3$ $T_{\rm o}$ decreases down to $T_{\rm
o}(x=1)\approx15\,{\rm K}$. This characteristic temperature is
obtained by scaling the $C_{\rm el}(T,x>0.3)$ results as a function
of the reduced temperature $T/T_{\rm o}$. At low Pd-content, the
antiferromagnetic ordering temperature $T_{\rm N}$ decreases very
rapidly from $T_{\rm N} = 36\,{\rm K}$ in pure CeRh$_2$Si$_2$ to
$T_{\rm N} = 18\,{\rm K}$ at $x = 0.1$. With higher Pd concentration
it stabilized at $T_{\rm N} \approx 15\,{\rm K}$ whereas the
magnitude of the anomalies in $C_{\rm el}(T)$ and in the
susceptibility around $T_{\rm N}$ are further reduced and disappear
at $x \approx 0.3$. This is in contrast to the behavior found on the
Pd-rich side, where $T_{\rm N}$ decreases continuously to zero with
increasing Rh content. The pronounced differences observed between
both phase boundaries and the drastic effect of doping on the Rh rich
side suggest an itinerant character in CeRh$_2$Si$_2$, in contrast to
the localized character of CePd$_2$Si$_2$. A further evidence for the
itinerant character in CeRh$_2$Si$_2$ is given by the $\rho(T)$
dependence observed for $x \le 0.3$, which scales with $\rho(T)$ of
the prototype itinerant compound YCo$_2$. The exponent ($n$) of the
power law observed in $\rho(T)$ at low $T$ presents a broad minimum
with $n \approx  1.2$ in the intermediate concentration range $0.4
\le x \le 0.6$, which we attribute to the effect of the random atomic
distribution of Rh and Pd in this region.
\PACS{{71.10Hf}{Non-Fermi-liquid ground states, electron phase
diagrams and phase transition in model systems} \and {71.27.+a}{
Strongly correlated electron systems; heavy fermions}}
}
\authorrunning{M. G\'omez Berisso et al.}
\titlerunning{Study of the Ce(Rh$_{1-x}$Pd$_x$)$_2$Si$_2$ alloy}
\maketitle

\section{Introduction}

The exceptionally high N\'eel temperature of CeRh$_2$Si$_2$, $T_{\rm
N}=36\,{\rm K}$,\cite{1Godart83} remains a puzzling question after almost
twenty years, having escaped the many attempts of explanation based on
mechanisms applicable to other Ce intermetalic compounds. The exceptional
magnitude of this $T_{\rm N}$ can be appreciated by comparing it with that of
GdRh$_2$Si$_2$ ($T_{\rm N}=90\,{\rm K}$) using the de\,Gennes
factor,\cite{2Szytula86} from which nearly two orders of magnitude difference
should be expected. The main difficulty for finding a realistic description of
this compound is related to the fact that it lies within the not yet well
defined boundary between the applicability of {\it local} or {\it itinerant}
models. In the case of CeRh$_2$Si$_2$, there are specific arguments in the
literature supporting each type of model. For the localized moment scenario,
one finds that: i) the magnetic moment along the $c$-axis\cite{3Quezel84}
($m_o=1.5 \mu_B$ at $T=T_{\rm N}/3$) is close to that of a Ce localized moment,
after correcting for crystal electric field effects (CEF); ii) the previously
reported specific heat jump\cite{4Graf98} at $T=T_{\rm N}$, $\Delta C_{\rm
el}(T_{\rm N})=10\,{\rm J/mol K}$, is close to the value for a local doublet
system in a mean field theory; iii) in coincidence, the related entropy gain
$\Delta S(T_{\rm N}) \approx R \ln 2$ is close to the value expected for a
doublet ground state (GS). Although a two step metamagnetic transition
(observed at 26\,Tesla) was taken as evidence for a local moment
system,\cite{5settai97} this argument is not conclusive because metamagnetic
transitions are also predicted for anisotropic itinerant systems,
\cite{6Gignoux97} a strong anisotropy being a clear property of the system at
hand.\cite{5settai97}

A strong mixing between the Ce-{\it 4f} and conduction states supporting the
itinerant description was recognized early from the large Curie-Weiss
temperature: \cite{3Quezel84} $\theta_W = -70\,{\rm K}$, suggesting a Kondo
temperature ($T_{\rm K}$) even larger than $T_{\rm N}$ for this compound,
though the strong anisotropy was observed in magnetic susceptibility ($\chi$)
on single crystals\cite{7araki98} with $\theta_{W \parallel c} \approx
-40\,{\rm K}$ (in the $c$ direction) and a value four times larger in the basal
plane. On the other hand, $T_{\rm K} \approx 33\,{\rm K}$ was obtained from the
width at the quasi-elastic line in inelastic neutron
scattering,\cite{Severing89} whereas NMR results suggest a much higher value
$T_{\rm K} \approx 100\,{\rm K}$.\cite{Kawasaki98} Neutron diffraction
experiments in CeRh$_2$Si$_2$ show an incommesurate antiferromagnetic structure
of itinerant character \cite{8kawarazaki95} close to $T_{\rm N}$, which becomes
commensurate at lower temperature. From the Si-NMR results much smaller
magnetic moment ($m_o = 0.22 \mu_B$) is estimated, and together with the high
$T_{\rm N}$ value suggests an itinerant character of the CeRh$_2$Si$_2$
magnetic ground state.\cite{9Delong91} The large difference in the $m_o$
determination between neutron diffraction and Si-NMR results indicates that the
characteristic time of the measurement is important due to the dynamic nature
of the magnetic correlations. Above $T_{\rm N}$, the electrical resistivity
($\rho$) increases on a characteristic energy scale of $T_{\rm o} \approx
200\,{\rm K}$, larger than $T_{\rm N}$ but comparable to an expected CEF
splitting.\cite{5settai97} These values, together with the $\rho(T)$ variation
under pressure supports the argument that band spin fluctuations contribute
significantly to the conduction electron scattering.\cite{10Thompson86}

Within the scope of a Doniach-type description,\cite{11Doniach77}
CeRh$_2$Si$_2$ is a paradigmatic case. With the highest $T_{\rm N}$ value among
the CeT$_2$Si$_2$ compounds it has to be placed at the top of the Doniach
diagram.\cite{12Endstra93} Such a position is coherent with the $T_{\rm N}$
evolution of the CeRh$_2$(Si,Ge)$_2$ system,\cite{13Godart89} where the
increase of the volume due to the substitution of Si by Ge leads to a decrease
of $T_{\rm N}$. This argument is complemented by the pressure dependence
measurements on stoichiometric CeRh$_2$Si$_2$, which also shows a decrease of
$T_{\rm N}$ down to about 10\,K at approximately 1\,GPa and the disappearance
of the related anomaly above that pressure.\cite{14Thompson00} However, when
the comparison is performed with respect to other CeT$_2$Si$_2$ compounds it
becomes contradictory in terms of the absolute $T_{\rm N}$ and $T_{\rm K}$
values, because for T=Rh the Kondo temperature is much larger than that for
T=Pd or Cu, despite the fact that these compounds have a ``less magnetic''
behavior.\cite{12Endstra93} The pressure effect shows that, despite the high
$T_{\rm N}$ value, the magnetic order breaks down at the lowest pressure value
within this family of Ce-compounds. The extreme sensitivity of this magnetic
interaction is also evident with respect to GdRh$_2$Si$_2$ (there 1GPa reduces
$T_{\rm N}$ by only 10\% \cite{2Szytula86}), or CePd$_2$Si$_2$ (which requires
a three times higher pressure to suppress the magnetic order despite the fact
that $T_{\rm N}$ is three times lower\cite{15Mathur98}).

At present, the large amount of information accumulated on CeRh$_2$Si$_2$ is
not conclusive enough to elucidate whether this compound has to be considered
as {\it local} or {\it itinerant} in its magnetic behavior. This ambiguity is
mainly related to the fact that it is placed at the peculiar position where the
energy associated to competing parameters, like $T_{\rm N}$ and $T_{\rm K}$,
are comparable. New independent information can be provided by studying the
evolution of this system when those parameters are continuously modified
driving the compound to a more {\it local} scenario. This purpose can be
achieved only by selective alloying, because pressure increases the itinerant
character by increasing the hybridization. The chance to enhance the local
character is provided by partial substitution of Rh with Pd in the
Ce(Rh$_{1-x}$Pd$_x$)$_2$Si$_2$ system, which was recently shown to form
continuously.\cite{16Trovarelli98} A preliminary investigation indicated a
complex magnetic phase diagram.\cite{17Berisso99} Taking advantage of the fact
that CePd$_2$Si$_2$ behaves as a localized magnet, a direct comparison of both
ends of the Ce(Rh,Pd)$_2$Si$_2$--phase-diagram should give more information
about the nature of the magnetic state of CeRh$_2$Si$_2$.

\section{Experimental and results}

For the present detailed study further samples on the Rh rich region and some
reference La(Rh$_{1-x}$Pd$_x$)$_2$Si$_2$ alloys were prepared following the
same sample preparation procedure and experimental techniques previously
described.\cite{16Trovarelli98,17Berisso99}

\begin{figure}
\begin{center}
\includegraphics[angle=0,scale=0.3]{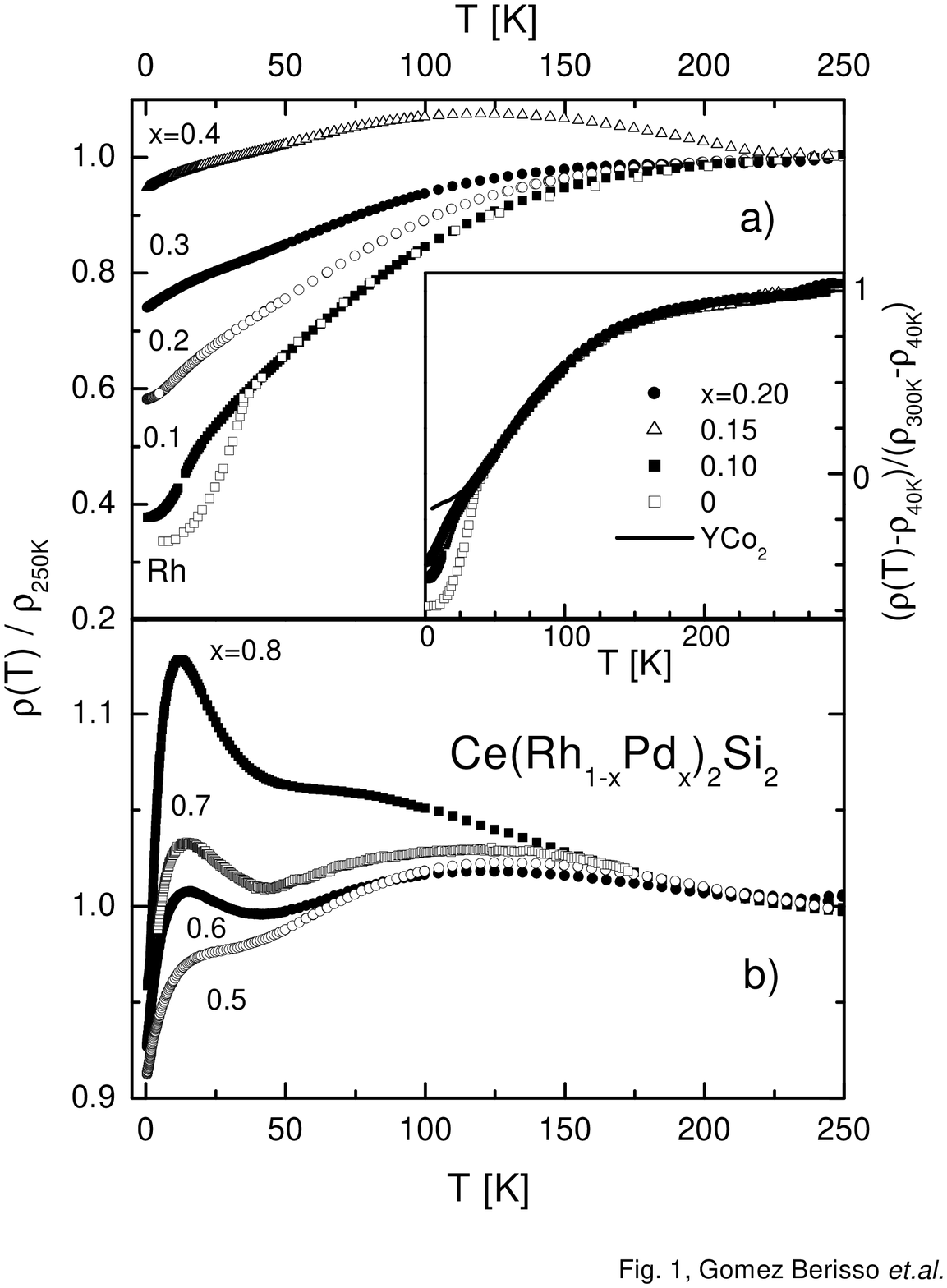}
\end{center}
\caption[]{Temperature dependence of the electrical resistivity
normalized to the value at 250\,K. a) Samples with $0 \le x \le 0.4$
and b) between $0.5 \le x \le 0.8$. Inset: scaling of the $\rho
(T,x)$ dependence for the Rh-rich samples with the $\rho (T)$
normalized at $T = 40\,{\rm K}$ and 300\,K. $\rho(T)$ of YCo$_2$
compound is also included for comparison. The data of the $x$=0
sample was taken from Ref.\cite{10Thompson86}.} \label{FigRho}
\end{figure}

Due to its sensitivity to the nature of the electronic scattering,
the electrical resistivity is one of the physical properties to be
investigated when a distinction between local and itinerant
electronic character is required. As reported in Fig.\,\ref{FigRho},
in this system the temperature dependence of the electrical
resistivity, $\rho (T)$, shows quite different features for both
concentration extremes. Due to the microscopical cracks in the
sample, the geometrical factor and consequently the absolute
resistivity values, cannot be determined unambiguously. Therefore the
$\rho (T)$ values were normalized at 250\,K for allowing a better
comparison. As already mentioned in Ref.\,\cite{5settai97}, in
stoichiometric CeRh$_2$Si$_2$ $\rho (T)$ increases continuously with
a characteristic energy scale of approximately 100\,K. This behavior
persists up to 30\% of Pd doping, but above that concentration
another relative maximum develops at low temperature (at approx.
$20\,{\rm K}$, as seen in Fig.\,\ref{FigRho}b). This maximum becomes
more pronounced with increasing Pd-content. The main feature is that
the temperature of both resistivity maxima practically does not
change with concentration, while the relative strength of the
electronic scattering changes significantly. On the Pd rich side, the
double maximum typical for trivalent Ce intermetallics compounds with
$T_{\rm K} \ll \Delta_{\rm CEF}$ (CEF-splitting) is observed (see
Fig.\,\ref{FigRho}b). The maximum at low temperature is attributed to
the electronic scattering by the GS and the other by the excited CEF
level, both being enhanced by the Kondo effect.\cite{21Cornut-72}

\begin{figure}
\begin{center}
\includegraphics[angle=0,scale=0.3]{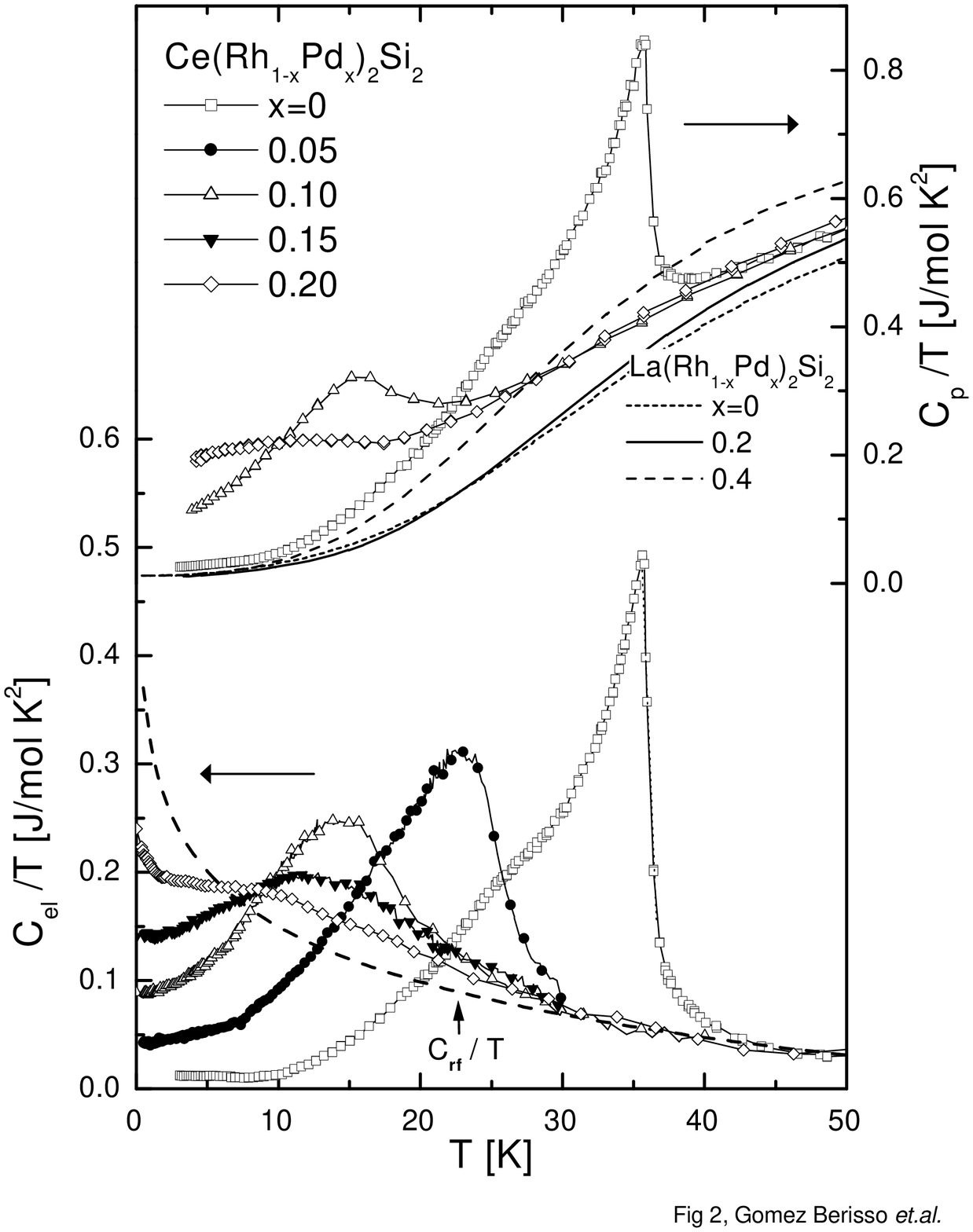}
\end{center}
\caption[]{Upper part: measured specific heat up to 50\,K of some Rh-rich and
(continuous curves) some La--reference alloys. Lower part: electronic
contribution to the specific heat of the Rh-rich samples ($0 \leq x \leq 0.2$).
The dash curve ($C_{\rm rf}/T$) is a function taken as reference for the
analysis of the entropy compensation (see the text).} \label{FigCp}
\end{figure}

The temperature dependence of the electronic contribution to the specific heat
($C_{\rm el}/T$) at low Pd doping ($0 \le x \le 0.2$) is shown in
Fig.\,\ref{FigCp}. This contribution was evaluated from the measured specific
heat ($C_{\rm p}/T$) as: $C_{\rm el}/T = C_{\rm p}/T-C_{\rm ph}/T$, where
$C_{\rm ph}/T$ is the phonon contribution extracted from La isotypic compounds
with $x$=0, 0.2, 0.4 and 1. As it can be seen in the upper part of
Fig.\,\ref{FigCp}, the Ce-based samples on the Rh-rich side ($0 \leq x \leq
0.2$) show quite similar $C_{\rm p}/T$ values at $T \geq 40\,{\rm K}$. Below
10\,K, La-based samples on the Rh rich side ($x$=0 and 0.2) also show a little
difference in the phonon contribution. In contrast, from $x=0.4$ to $x=1$,
$C_{\rm ph}(T)/T$ increases significantly with $x$ after a drastic enhancement
between $x=0.2$ and 0.4. Therefore the phonon subtraction for the Ce-based
samples was calculated using  the extrapolation of the La $x=0$ and 0.2 samples
for the Rh rich region and that between $x=0.4$ and 1 for the rest of the
samples. This non-monotonous variation of $C_{\rm ph}$ upon Pd doping suggests
a change in the phonon spectrum around $x=0.3$. However, no structural
transition was detected from the X-ray data: only a change in the
$x$-dependence of the $c/a$-ratio at that concentration.

Our sample of pure CeRh$_2$Si$_2$ shows a larger jump at $T_{\rm N}$, $\Delta
C_{\rm el}(T_{\rm N}) = 15\,{\rm J/mol\,K^2}$, than previously reported in the
literature,\cite{4Graf98} which exceeds the mean field prediction. This peak in
$C_{\rm el}/T$ at $T_{\rm N}$ is better observed in thermal
expansion,\cite{7araki98} and is probably related to the opening of a magnetic
excitation gap due to the strong Ising-type anisotropy of this compound. A
further weak anomaly associated with the change of the magnetic propagation
vector is also observed at 25\,K.\cite{22Grier84}

\begin{figure}
\begin{center}
\includegraphics[angle=-90,scale=0.3]{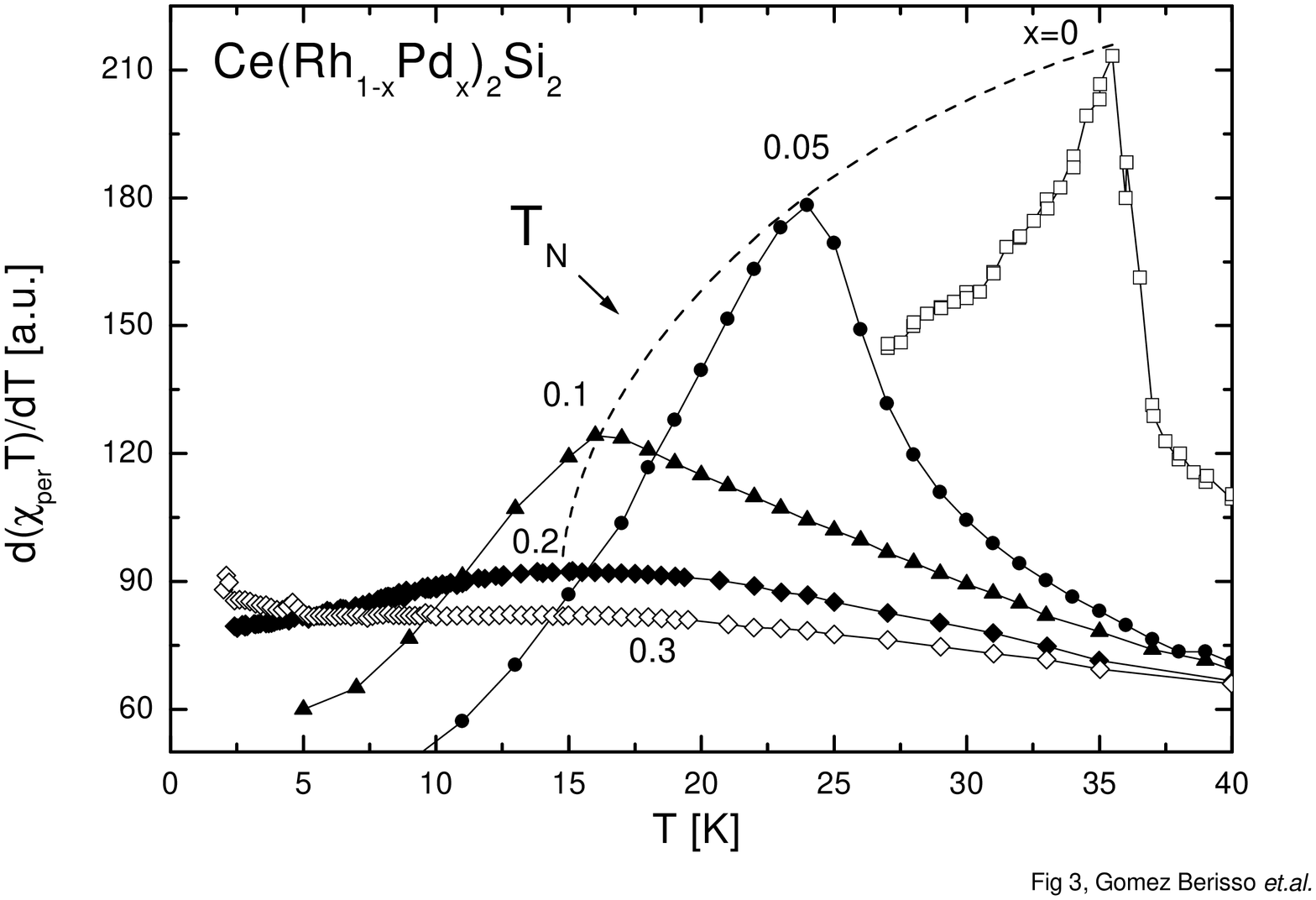}
\end{center}
\caption[]{Plot of the derivative of $\chi\,T$ as a function of the
temperature and the concentration in order to better determine
$T_{\rm N}(x)$ on the Rh-rich side. The magnetic response was taken
from Ref.\cite{16Trovarelli98} with the field perpendicular to the
textures.} \label{FigChi}
\end{figure}

A small amount of Pd ($x \le 0.1$) already leads to a strong decrease of
$T_{\rm N}$ and a pronounced broadening of the anomaly in $C_{\rm el}(T)$.
Further increase of the Pd concentration ($0.1 \le  x  \le 0.4$) leaves the
temperature of the maximum almost unchanged but reduces the size of the
anomaly, which eventually disappears between $x = 0.3$ and $x = 0.4$. Due to
the broadening, the analysis of the anomaly in $C_{\rm el}(T)$ does not lead to
a reliable determination of $T_{\rm N}(x)$. More precise values can be obtained
from the susceptibility by looking at the derivative $d(\chi\,T)/dT$ and
defining $T_{\rm N}$ as the temperature of the maximum (see
Fig.\,\ref{FigChi}). This demonstrates very clearly the rapid drop of $T_{\rm
N}$ from 36\,K in pure CeRh$_2$Si$_2$ to 18\,K at $x = 0.1$ and then the
leveling off at around $15\,{\rm K}$ for $x \le 0.2$ (see also the phase
diagram in Fig.\,\ref{FigDiag}). This is in sharp contrast to the Pd-rich
region, where $T_{\rm N}$ drops monotonously with increasing Rh content and
extrapolates to 0\,K at $x = 0.65$. The broadening of the anomaly is also much
less pronounced on the Pd-rich side. The disappearance of the anomaly without a
concomitant decrease of $T_{\rm N}$ to 0\,K on the Rh-rich side indicates that
the degrees of freedom involved in the magnetic transition as well as the free
energy gained in that transition decrease with Pd content and eventually
vanish, whereas $T_{\rm N}$ and thus the strength of the magnetic interaction
still has a finite value.

Within the intermediate concentration region ($0.3 < x < 0.7$), $C_{\rm
el}(T)/T$ is well described by a logarithmic decrease, with a downward
curvature at low temperature, as typically observed in systems lying close to a
magnetic instability.\cite{20Moriya} The low temperature $C_{\rm el}/T$ value
increases proportionally to the Pd-concentration up to $x = 0.7$, where the
onset of the magnetic order of CePd$_2$Si$_2$ is observed.

\section{Discussion}

\begin{figure}
\begin{center}
\includegraphics[angle=0,scale=0.3]{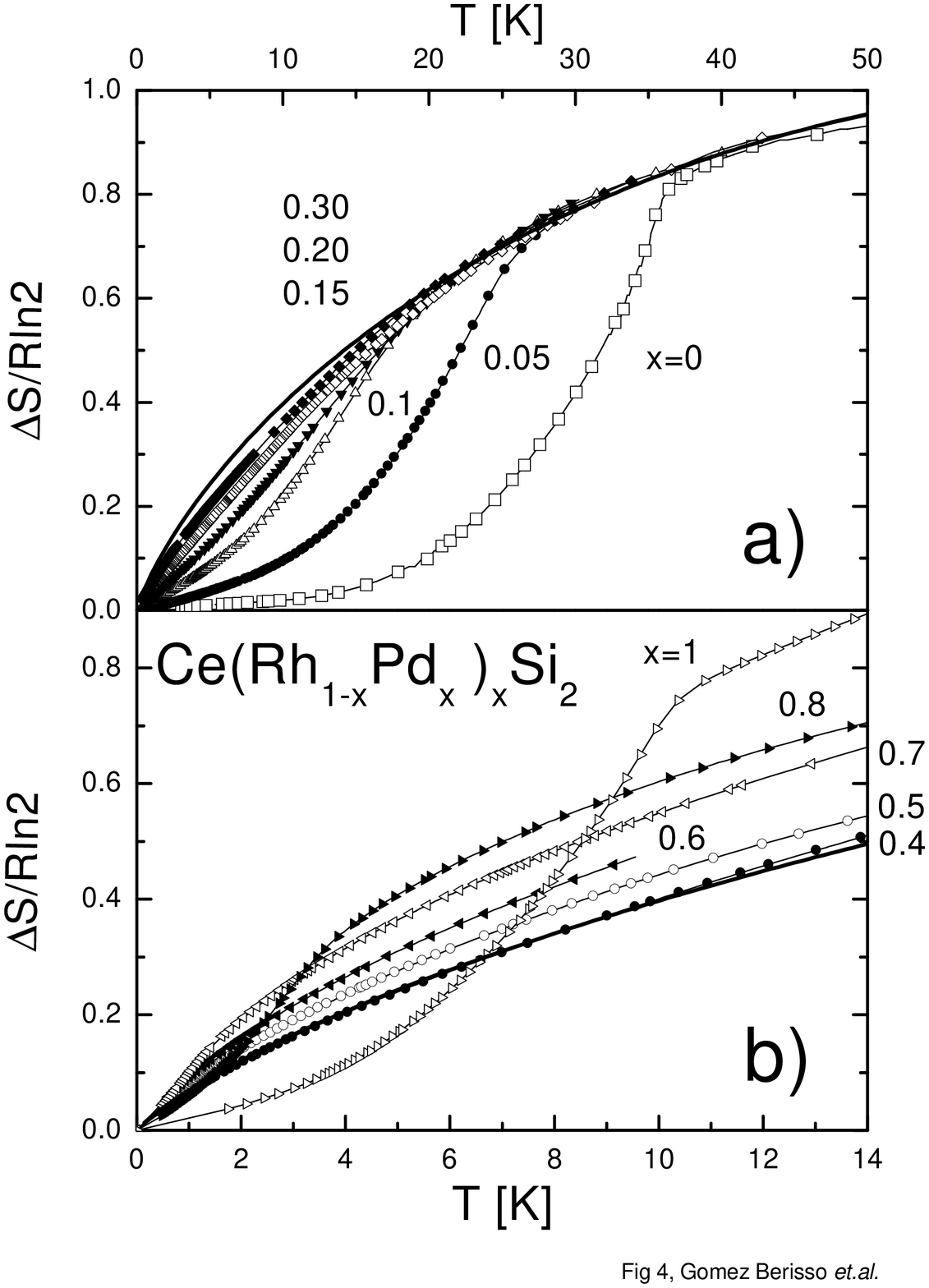}
\end{center}
\caption[]{Evolution of the temperature dependence of the magnetic entropy as a
function of Pd concentration. a) low Pd content and b) intermediate and high Pd
concentration. The continuous curve represents the entropy related to the
reference curve $C_{\rm rf}/T$ proposed in Fig.\,\ref{FigCp}.} \label{FigEntro}
\end{figure}

Once the intrinsic differences of doping effect on $T_{\rm N}$ was establishes
between Pd and Rh rich sides, we shall analyze further properties to gain
insight into the nature of the ground state on both extremes of the alloy
system. The most important parameter is the characteristic energy related to
the delocalization of the 4{\it f}-electrons. A rough idea about its dependence
on the composition can be obtained by looking at the evolution of the entropy.
In Fig.\,\ref{FigEntro}a, we show the entropy gain as a function of temperature
up to 50\,K on the Rh-rich side, and in Fig.\,\ref{FigEntro}b that of the
intermediate and Pd-rich samples up to 14\,K. Within the experimental
dispersion, the entropy gains in the Rh-rich samples merge above the respective
$T_{\rm N}(x)$ into a common curve. As it will be shown later, this curve can
be described by a simple function which relates the results in the Rh-rich
samples to those at intermediate and high Pd-contents. The merging to a common
function means that one can define a magnetically-non-ordered state (hereafter,
normal state) for all the Rh-rich samples. The temperature dependence of the
(electronic/magnetic) entropy $\Delta S(T)$ of this normal state corresponds to
the common function and determines the entropy observed at $T_{\rm N}(x)$ for a
given composition. This is a result expected in the case where one energy scale
determines $\Delta S(T)$ (and thus $C_{\rm el}(T)/T$ or $C_{\rm rf}/T$ as in
Fig.\,\ref{FigCp}b) of the normal state whereas a second independent energy
scale determines the ordering temperature. This can be compared with some quasi
one-dimensional spin systems, where $C_{\rm el}(T)/T$ of the normal state is
determined by the {\it intrachain} exchange and $T_{\rm N}$ by the {\it
interchain} exchange, or for a superconductor (or a Spin Density Wave) were the
normal state $C_{\rm el}(T)/T$ is determined by the electron density of states
(i.e., the band width) whereas $T_{\rm C}$ (or $T_{\rm SDW}$) is determined by
the interaction between the quasiparticles. In those cases, the change of the
ordering temperature $T_{\rm C}$ ($T_{\rm SDW}$) leads to a change of the
entropy at $T_{\rm C}$ ($T_{\rm SDW}$) in accord with the temperature
dependence of the entropy of the normal state. However, this would not be the
case in a {\it purely} localized three-dimensional antiferromagnet where,
reducing the exchange strength (and thus $T_{\rm N}$) would not change the
entropy collected at $T_{\rm N}$. Therefore this result is a very strong
indication that in the Rh-rich region there is a characteristic {\it 4f}-energy
which is independent of the composition.  This characteristic energy governs
the normal state and it is not related to the magnetic order but to the
hybridization energy of the {\it 4f}-electrons.

\begin{figure}
\begin{center}
\includegraphics[angle=-90,scale=0.3]{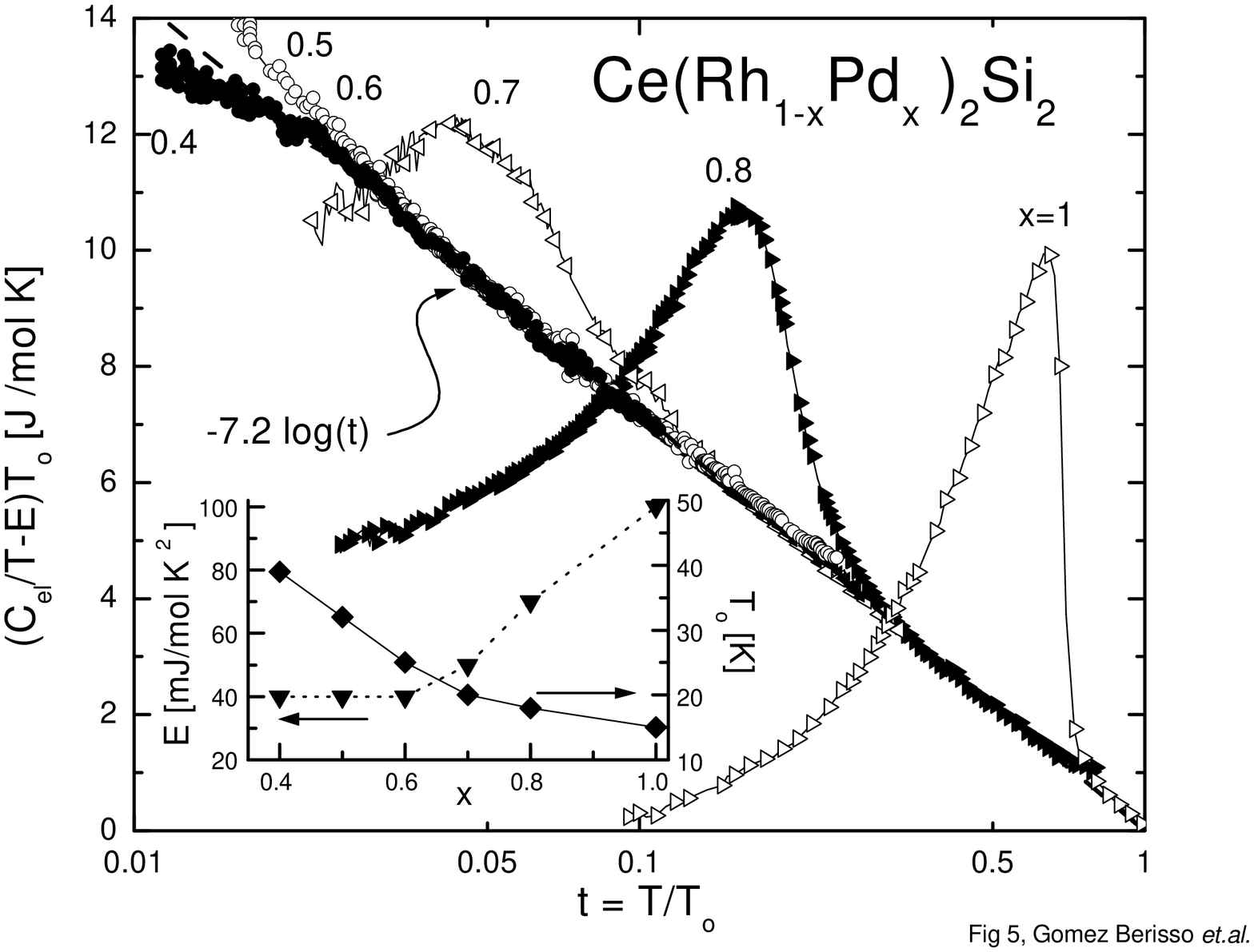}
\end{center}
\caption[]{Scaling of the specific heat as a function of a normalized
temperature $t=T/T_{\rm o}$. Inset: Evolution of the two fitting parameters
$T_{\rm o}$ and $E$ as a function of Pd doping in the $0.4 \leq x \leq 1$
range.} \label{FigScal}
\end{figure}

In contrast to the Rh-rich region, the entropy gain in the intermediate and in
the Pd-rich regions increases continuously with the Pd-content. This indicates
a continuous decrease of the characteristic energy with increasing $x$ for $x
\ge 0.4$. A more precise analysis can be performed by fitting the $C_{\rm
el}(T)/T$ data of the samples with $0.4 \ge x \ge 1$ using a scaling formula
proposed for systems close to a magnetic instability, \cite{26Sereni} $C_{\rm
el}/t = -7.2 \log(t) + E\,T_{\rm o}$ with $t = T/T_{\rm o}$, $T_{\rm o}$ and
$E$ being two free parameters corresponding to a characteristic energy and the
linear background contribution to the specific heat, respectively. The fits and
the concentration dependence of $T_{\rm o}$ and $E$ are shown in
Fig.\,\ref{FigScal}. $T_{\rm o}$ decreases continuously from 40\,K at $x = 0.4$
to 15\,K at $x = 1.0$, whereas $E$ increases continuously from 40 to
100\,mJ/mol\,K$^2$. The corresponding $\Delta S(T)$ and $C_{\rm el}(T)/T$
curves are shown as reference functions in Fig.\,\ref{FigEntro}b (continuos
line) and Fig.\,\ref{FigScal} (straight line), respectively. The fitting
parameters, obtained for $x$=0.4, can be used to fit the entropy of the common
normal state observed in the Rh-rich region indicating that, despite the change
of regime, there is a smooth evolution of the normal state from the $x < 0.4$
to the $x > 0.4$ region. One should notice that the $T_{\rm o}$ value we obtain
for CeRh$_2$Si$_2$ and CePd$_2$Si$_2$ (42\,K and 15\,K, respectively) are close
to the $T_{\rm K}$ values of 33\,K and 10\,K given by the inelastic neutron
scattering.\cite{Severing89} This supports the applicability of the scaling
formula. The scaling further implies that, in the absence of magnetic order
(i.e., in the concentration range $0.3 < x < 0.7$), $C_{\rm el}/T$ is inversely
proportional to $T_{\rm o}$ at very low temperatures, as it occurs in the
single ion Kondo model with $T_{\rm K}$. For the samples showing magnetic order
this is no longer true, since part of the degrees of freedom contributing to
$C_{\rm el}/T$ condense into the magnetic state. Thus the analysis of the
composition dependence of $\Delta S(T)$ and $C_{\rm el}(T)/T$ demonstrates a
continuous evolution of the normal state from pure CeRh$_2$Si$_2$ to pure
CePd$_2$Si$_2$. Nevertheless, a clear break in the $x$-dependence of the
characteristic energy $T_{\rm o}$ occurs at $x = 0.4$, since $T_{\rm o}$ is
constant for $x < 0.4$ but decreases continuously for $x \ge 0.4$. This change
of regime underscores the different nature of the Ce ground state in both
concentration limits.

As mentioned before, no discontinuity in the crystalline parameters is observed
in this system\cite{16Trovarelli98} the ``$c/a$''-ratio, as was already
mentioned, undergoes one of the largest variations observed among the {Ce\,122}
intermetallics with ThCr$_2$Si$_2$ type structure,\cite{Just96} changing from
$c/a =2.49 $ for CeRh$_2$Si$_2$ to 2.33 for CePd$_2$Si$_2$, with a weak
variation in the slope at $x=0.3$. A concomitant modification in the magnetic
structure is observed between CeRh$_2$Si$_2$ (with the moments ordered along
the $c$-axis) and CePd$_2$Si$_2$ (with the staggered magnetic moments on the
basal plane). \cite{22Grier84} Such a drastic difference makes a continuity in
the LRMO between both stoichiometric extremes unlikely. One would instead
expect a disordered or frustrated magnetic region between those phases, but our
results show that also the character of the {\it f}-electron localization is
changing.

Further information can be obtained from a more detailed analysis of the
resistivity, since its temperature dependence is dominated by the effect of the
magnetic scattering. On the Rh-rich side, a clear scaling in the $\rho (T,x)$
dependence can be observed when the measured $\rho (T)$ is normalized to its
value at 40\,K right above $T_{\rm N}(x=0)$ and at high temperature, i.e.
$(\rho(T)-\rho_{40\,{\rm K}})/(\rho_{300\,{\rm K}}- \rho_{40\,{\rm K}})$, as
displayed in the inset of Fig.\,\ref{FigRho}a. This scaling holds only for the
alloys that show magnetic order (i.e. $x \leq 0.2$), indicating that the
magnetic component involved in the electronic scattering has a different nature
for the low Pd doping range than in the intermediate region. It is worth noting
that the resistivity of YCo$_2$,\cite{24Hauser98} a prototype of band spin
fluctuation system, fits into this scaling, as shown in the inset of
Fig.\,\ref{FigRho}a. This is a strong evidence for the itinerant character of
the electronic properties of these alloys. Such a scaling also supports the
fact that the energy which characterizes the Rh-rich region does not change
with concentration. Despite its relatively large curvature, the $\rho(T)$
dependence cannot be attributed to an intermediate valence behavior (like in
CeRh$_2$ for example\cite{25Harrus85}) because of LRMO at lower temperature.

\begin{figure}
\begin{center}
\includegraphics[angle=-90,scale=0.3]{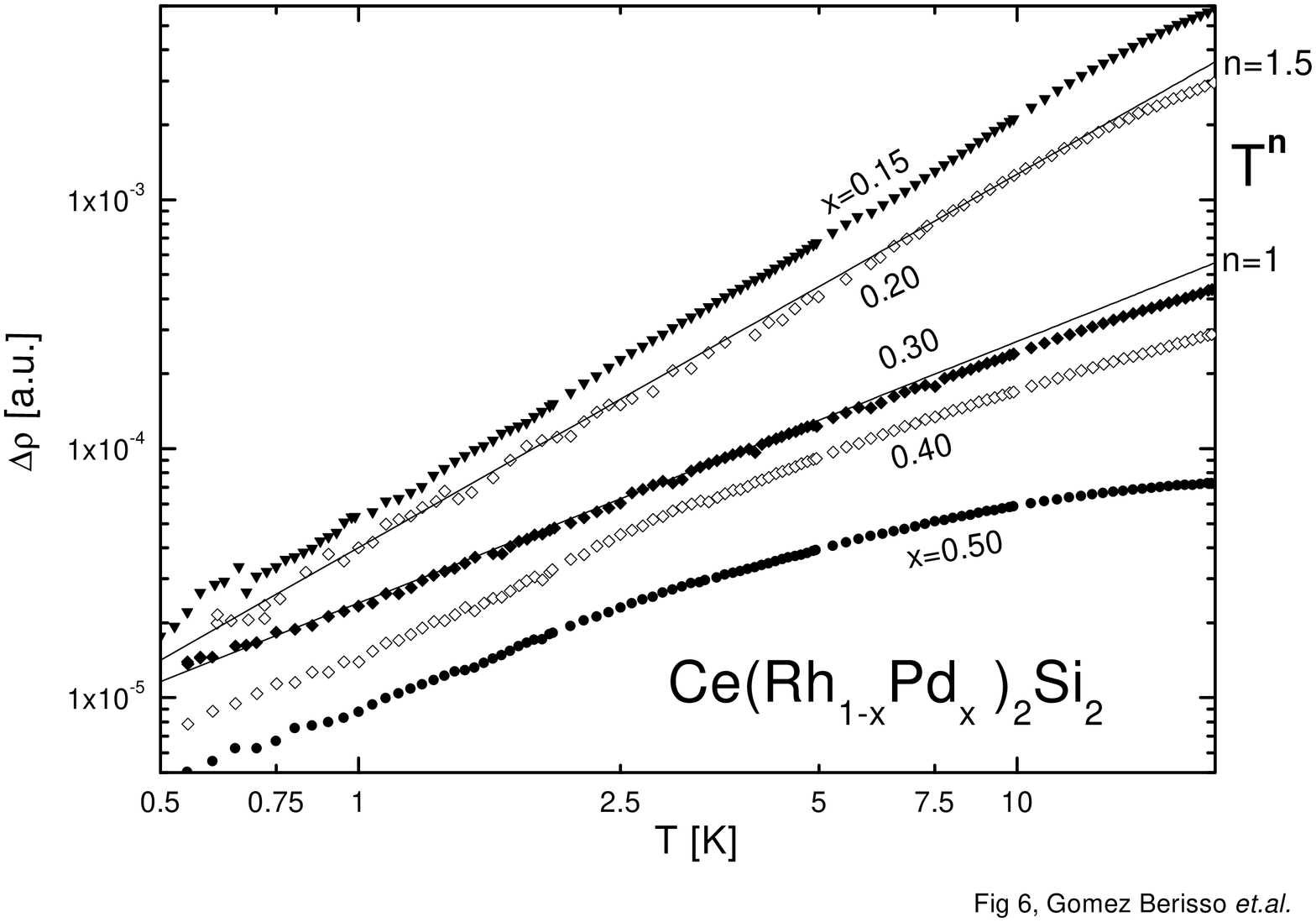}
\end{center}
\caption[]{Power law temperature dependence  of the magnetic component, $\Delta
\rho (T) \propto T^n$, of the electrical resistance in the low temperature
region in a double logarithmic representation for the $0.15 \leq x \leq 0.5$
range.} \label{FigN}
\end{figure}

\begin{figure}
\begin{center}
\includegraphics[angle=-90,scale=0.3]{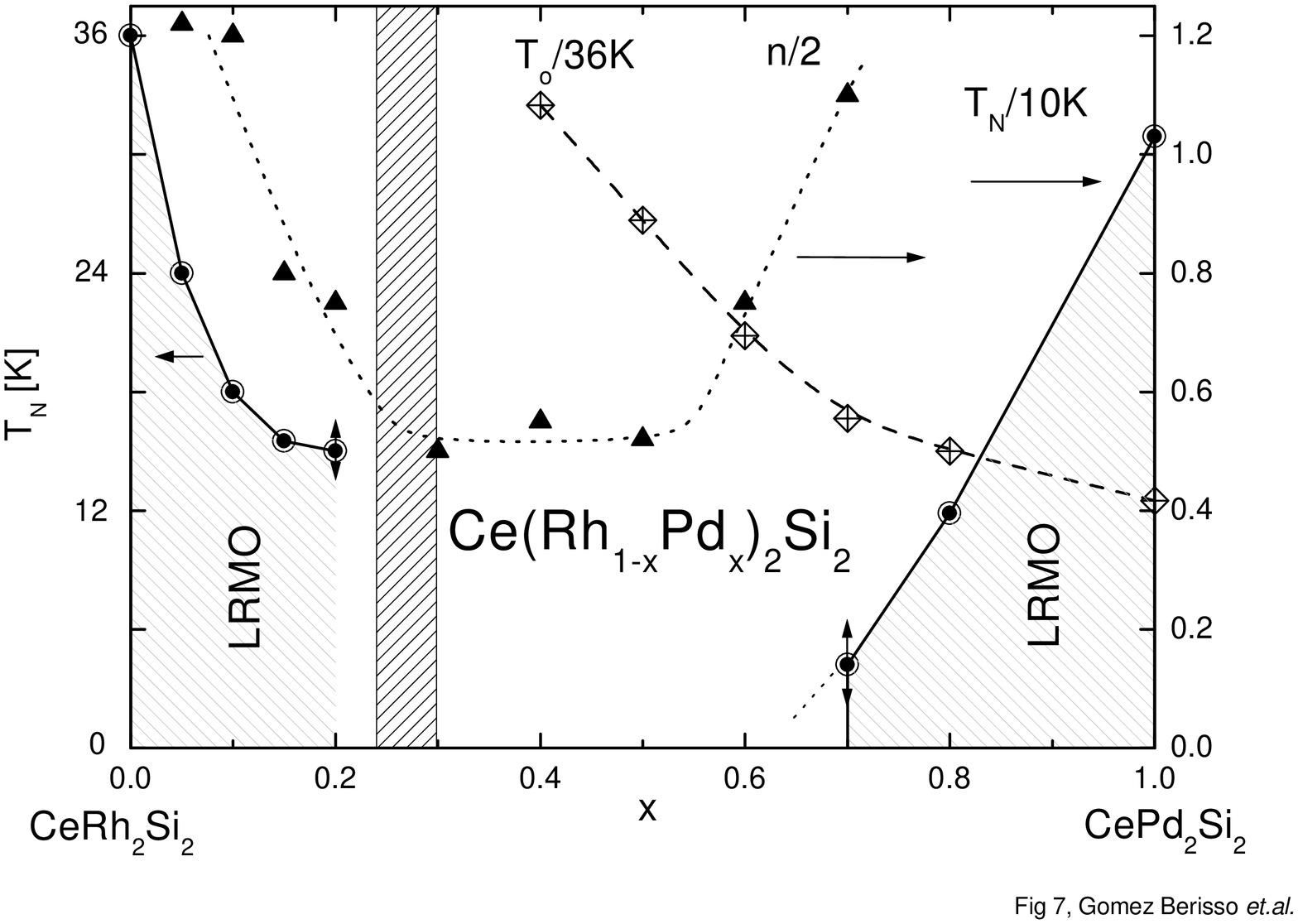}
\end{center}
\caption[]{Magnetic phase diagram showing as a function of Pd doping ($x$), the
existence region of the magnetic ordered phases (LRMO), the transition region
(shadowed area), the evolution of the normalized scaling factor $T_{\rm o}(0.4
\leq x \leq 1)$ and the exponent of the power law dependence of $\Delta \rho
(T) \propto T^n$ are also shown.} \label{FigDiag}
\end{figure}

The disappearance of the magnetic order is connected with a profound change in
the low energy excitations. This is evidenced in the resistivity, which at low
temperatures was found to follow a power law $\rho(T) = \rho_0 + a\,T^n$ with
an exponent $n$ that changes systematically with composition. As an example we
show in Fig.\,\ref{FigN} the temperature dependent part of the resistivity,
$\Delta \rho (T) = \rho (T) - \rho_0$, in a log-log plot for the concentration
range $0.15 \le x \le 0.5$. Here, the disappearance of the magnetic order leads
to a strong decrease of $n$ to values close to 1, characteristic for non-Fermi
Liquid systems. \cite{28hvl} For $x < 0.2$ and $x > 0.6$ we found values of $n$
larger than 2 (see Fig.\,\ref{FigDiag})as expected for compounds showing LRMO.

The present investigation of the Ce(Rh$_{1-x}$Pd$_x$)$_2$Si$_2$ system allows
us to propose a more detailed phase diagram. We have included in
Fig.\,\ref{FigDiag} the magnetic phase boundaries, $T_ {\rm N}(x)$, the
evolution of the characteristic energy $T_{\rm o}$ as deduced from the scaling
of $C_{\rm el}(T)/T$, as well as the exponent $n$ of the power law in $\rho(T)$
at low temperatures. The change in the evolution of $T_{\rm o}$, as deduced
from the $C_{\rm el}(T)/T$ and $\rho(T)$ results, suggests the division of the
magnetic phase diagram into two regions with a crossover region between $x =
0.3$ and $x = 0.4$ (hatched area). On the Rh-rich side, a strong decrease of
$T_{\rm N}(x)$ coexist with a constant $T_{\rm o}$, while the suppression of
the magnetic order occurs at a finite $T_{\rm N}$ in the crossover region. On
the Pd-rich side of the crossover region, $T_{\rm o}$ decreases continuously
with increasing $x$, with the slope becoming  weaker once the magnetic order
appears. The part of the phase diagram on the right (Pd-rich) side of the
crossover region corresponds to the expected for a transition from a
non-ordered to a magnetically ordered Kondo-lattice system. Simple appropriate
models, like that proposed by Doniach, predict a monotonous decrease of $T_{\rm
N}$ with increasing $T_{\rm K}$ in the vicinity of the transition, just as
observed here for increasing Rh doping in CePd$_2$Si$_2$. Therefore this part
of the phase diagram is in agreement with the present picture of CePd$_2$Si$_2$
as a localized antiferromagnet. In contrast, the part of the phase diagram on
the left (Rh-rich) side of the crossover region cannot be explained within such
a localized Doniach-model, since the magnetic order is suppressed without a
concomitant increase of the characteristic energy. There, the scaling of the
resitivity with that of YCo$_2$ and the evolution of the entropy with doping
support an itinerant type of magnetic order. Itinerant magnets are sensitive to
disorder (more than localized ones), especially if the ordered state is close
to the stability limit. This is certainly the case for CeRh$_2$Si$_2$ as shown
by the strong suppression of the LRMO with pressure (one order of magnitude
larger than for CePd$_2$Si$_2$). Then, the rapid suppression of the magnetic
order upon Pd-doping despite a constant characteristic energy can easily be
accounted for by the disorder due to the doping.

\section{Conclusions}

The main result of this investigation is that the Rh-rich and the Pd-rich parts
of the Ce(Rh$_{1-x}$Pd$_x$)$_2$Si$_2$ system behave very differently. On the
Pd-rich side, increasing the Rh-content leads to a pronounced increase of the
characteristic {\it 4f}-energy $T_{\rm o}$ and a concomitant continuous
decrease of $T_{\rm N}$ down to 0\,K. This side of the magnetic phase diagram
corresponds to the predictions of models based on localized {\it f}-electrons,
supporting the current interpretation of a localized antiferromagnetic state in
CePd$_2$Si$_2$. In contrast, on the Rh-rich side, Pd doping leads to a
continuous decrease of the degrees of freedom involved in the magnetic ordered
state and to the disappearance of the magnetic state at a finite $T_{\rm N}$,
despite the fact that the characteristic energy $T_{\rm o}$ is not affected by
the Pd-doping. This points to an itinerant type of magnetic order in
CeRh$_2$Si$_2$, which is destroyed by the disorder introduced by the Pd-doping.
Further evidence for the itinerant character is that $\rho (T)$ in the Rh-rich
region scales with that of YCo$_2$, a prototype band spin-fluctuation system.
The change of slope in the dependence of  $T_{\rm o}$ on composition between $x
= 0.3$ and $x = 0.4$ suggests that the change from the itinerant to the
localized regimes takes place in that region. In the intermediate region $0.4 <
x < 0.6$, we observe in $\rho (T<{5\,K})$ a power law with an exponent $n$
close to 1, which we attribute to the disorder induced by the alloying and to
the absence of long range magnetic order.

\begin{acknowledgement}

This work was partially supported by joint programs between Fundacion
Antorchas (Arg.) and Alexander von Humboldt Stiftung (Germ.)
\#A-13391/1-003, Fundaci\'on Antorchas and DAAD (Germ.) \#13740/1-88
and ANPCyT: project PICT \#03-3250. MGB, PP and JGS, are members of
the Consejo Nacional de Investigaciones Cient\'{\i}ficas y T\'ecnicas
(Arg.).
\end{acknowledgement}

\end{document}